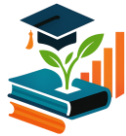
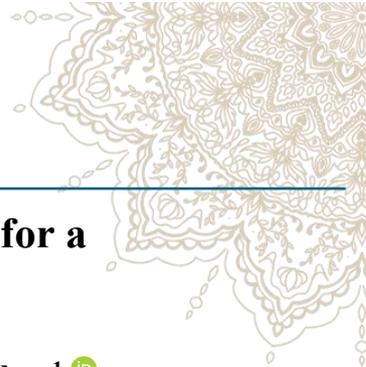

## Education and Development Lab
**Education Research Team [Education and Computer Science]**

# Green Computing: The Ultimate Carbon Destroyer for a Sustainable Future

**Sayed Mahbub Hasan Amiri[1,*]**  , **Prasun Goswami[2]**  , **Md. Mainul Islam[1]**  ,

**Mohammad Shakhawat Hossen[3]**  , **Marzana Mithila[4]** , **Naznin Akter[5]**

[1] Department of ICT, Dhaka Residential Model College, Bangladesh, [2] Department of English, Dhaka Residential Model College, Bangladesh, [3]Department of ICT, Char Adarsha College, Kishoreganj, Bangladesh, [4]Field Work Department, Unique Personnel (UK) Limited, London, UK, [5]Department of English, Shamplapur Ideal Academy, Bangladesh.

**Abstract**

Green computing represents a critical pathway to decarbonize the digital economy while maintaining technological progress. This article examines how sustainable IT strategies including energy-efficient hardware, AI-optimized data centres, and circular e-waste systems can transform computing into a net carbon sink. Through analysis of industry best practices and emerging technologies like quantum computing and biodegradable electronics, we demonstrate achievable reductions of 40-60% in energy consumption without compromising performance. The study highlights three key findings: (1) current solutions already deliver both environmental and economic benefits, with typical payback periods of 3-5 years; (2) systemic barriers including cost premiums and policy fragmentation require coordinated action; and (3) next-generation innovations promise order-of-magnitude improvements in efficiency. We present a practical framework for stakeholders from corporations adopting renewable-powered cloud services to individuals extending device lifespans to accelerate the transition. The research underscores computing's unique potential as a climate solution through its rapid innovation cycles and measurable impacts, concluding that strategic investments in green IT today can yield disproportionate sustainability dividends across all sectors tomorrow. This work provides both a compelling case for urgent action and a clear roadmap to realize computing's potential as a powerful carbon destruction tool in the climate crisis era.

*Corresponding author: Sayed Mahbub Hasan Amiri

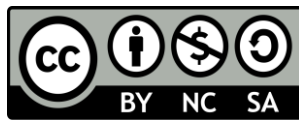

## 1. Introduction

The 21st century has witnessed unprecedented environmental challenges, with climate change emerging as the defining crisis of our era. The planet's average temperature has risen by 1.1°C since the pre-industrial period, primarily due to greenhouse gas (GHG) emissions, with carbon dioxide ($CO_2$) being the most significant contributor (IPCC, 2021). The energy sector remains the largest emitter, accounting for 73% of global $CO_2$ emissions (IEA, 2022). However, as industries shift toward renewable energy, another major polluter is rapidly growing the Information Technology (IT) sector. From massive data centres humming 24/7 to billions of electronic devices discarded annually, the digital revolution has come at an environmental cost. If the Internet were a country, it would rank as the world's fourth-largest polluter, surpassing entire nations in electricity consumption (The Shift Project, 2020). This raises a critical question: Can technology, which has accelerated climate change, also become its solution?

Despite its reputation as a "clean" industry, the Information Technology (IT) sector harbours a substantial and often overlooked environmental impact. Data centres alone consume about 1% of global electricity and are responsible for approximately 2% of total $CO_2$ emissions comparable to the aviation industry (Nature, 2021; Uptime Institute, 2023). Meanwhile, the explosion of electronic waste adds to the crisis, with over 53 million metric tons of e-waste generated globally in 2023, yet only 17% is properly recycled, leaving toxic substances to contaminate soil and water systems (Global E-Waste Monitor; UNEP, 2023). Additionally, the manufacturing and supply chain processes in IT are major emission sources; for instance, producing a single laptop can release between 300 to 500 kg of $CO_2$, and semiconductor fabrication demands immense quantities of water and energy (Carbon Trust). As demand for technologies like cloud computing, artificial intelligence, and the Internet of Things continues to surge, these environmental challenges are set to intensify unless the industry transitions toward sustainable and green computing practices.

Although commonly viewed as a low-impact or "clean" industry, the IT sector significantly contributes to global environmental degradation through its high energy use, waste production, and carbon emissions. Energy-hungry data centres are a major concern, consuming around 1% of the world's electricity and accounting for nearly 2% of global $CO_2$ emissions equivalent to the aviation industry (Nature, 2021; Uptime Institute, 2023). The rapid turnover of electronic devices has also led to a surge in e-waste, with over 53 million metric tons generated in 2023 alone. Shockingly, only 17% of this waste is properly recycled, leaving hazardous substances to seep into ecosystems (Global E-Waste Monitor; UNEP, 2023). Moreover, the manufacturing process itself is carbon-intensive: producing a single laptop emits between 300 and 500 kilograms of $CO_2$, and semiconductor fabrication requires enormous amounts of water and energy (Carbon Trust). As the global appetite for cloud computing, artificial intelligence, and Internet of Things (IoT) technologies continues to rise, these environmental challenges will only escalate unless the industry commits to greener, more sustainable computing practices.

Green Computing, also known as Sustainable IT, encompasses the environmentally responsible design, production, use, and disposal of computers, servers, and related technologies. Its core aim is to minimize the ecological footprint of the IT sector. Key principles include enhancing energy efficiency through optimized hardware and software, powering operations with renewable energy sources such as solar, wind, or hydro, and reducing electronic waste by promoting device repair, reuse, and recycling. Additionally, technologies like virtualization and cloud optimization help decrease reliance on physical infrastructure by enabling shared computing resources. The concept of green computing gained initial traction in the 1990s with the launch of the Energy Star program, but the urgency surrounding climate change has since elevated its importance. Today, green computing is not only an environmental necessity but also a booming industry now valued at over $500 billion globally fuelled by increasing corporate sustainability commitments and evolving government regulations (Gartner, 2023).

With the IT sector's carbon footprint expected to double by 2025 (European Commission, 2022), green computing has evolved from a commendable initiative into a critical business and environmental necessity. One of the most pressing reasons is its role in combating climate change: although a single Google search emits just 0.2 grams of $CO_2$, the cumulative impact of 8.5 billion daily searches becomes significant (Stanford University). Transitioning to carbon-neutral data centers, like those pioneered by Google, has the potential to reduce global emissions by as much as 0.3 gigatons annually (Science Journal, 2022). Beyond the environmental gains, green computing also delivers strong economic advantages. Energy-efficient data centers save businesses approximately $30 billion each year in operational costs (McKinsey, 2023), while tech giants like Apple and Microsoft have reported up to 20% higher investor engagement due to their sustainability efforts (Bloomberg ESG Report). Furthermore, growing regulatory pressure adds urgency, with the European Union's Green Digital Pact mandating carbon-neutral IT infrastructure by 2030, and U.S. legislation like California's SB-343 enforcing strict e-waste management standards. In this landscape, adopting green computing is no longer optional it is essential for climate responsibility, cost efficiency, and long-term business resilience.

This article delves into how green computing acts as a powerful "carbon destroyer" by transforming the environmental impact of the IT sector through innovative, sustainable technologies. It begins by analysing energy-efficient advancements, such as AI-driven cooling systems like Google's DeepMind, which dramatically reduce data centre emissions, and the transition to ARM-based processors like Apple's M1 chip, which cuts power consumption by up to 60%. The article also examines how the IT industry is embracing renewable energy, highlighting pioneering efforts such as Microsoft's hydrogen-powered data centres and Facebook's use of wind energy to power its servers. In addressing the escalating e-waste crisis, it explores circular economy models, including Dell's closed-loop recycling system that recovers over 50 million pounds of e-waste annually, and Fairphone's modular smartphones designed to last over five years. Finally, the article looks ahead to future innovations, such as quantum computing for ultra-low-energy processing and biodegradable electronics designed to eliminate toxic waste. Together, these developments underscore the urgent and transformative role of green computing in reducing the carbon footprint of digital technologies.

The IT sector stands at a crossroads: continue an unsustainable path or lead the green tech revolution. With 60% of global GDP now digitized (World Bank), the industry's choices will shape our planet's future.

This article serves as a roadmap for businesses, policymakers, and consumers to embrace green computing not just as a trend, but as the ultimate carbon destroyer for a sustainable future.

## 2. Literature Review

### Historical Background of Green Computing

The concept of green computing emerged in response to growing concerns about energy consumption and environmental degradation caused by information technology. The movement gained formal recognition in 1992 with the launch of the Energy Star program by the U.S. Environmental Protection Agency (EPA), which set energy efficiency standards for computers and monitors (EPA, 2021). This initiative marked the first major effort to reduce IT-related power consumption, encouraging manufacturers to design devices that automatically entered low-power states when inactive.

A significant milestone occurred in 2007 with the formation of The Green Grid Consortium, an industry group that developed standardized metrics for measuring data center efficiency. The consortium introduced Power Usage Effectiveness (PUE), a ratio comparing total facility energy to IT equipment energy, which became a benchmark for sustainable data centre operations (The Green Grid, 2008). A PUE of 1.0 indicates perfect efficiency, while the global average in 2023 was 1.55, showing room for improvement (Uptime Institute, 2023).

Recent advancements have shifted toward AI-driven energy optimization. Google's DeepMind project demonstrated how machine learning could reduce data centre cooling energy by 40% through predictive algorithms (Evans & Gao, 2023). This breakthrough highlighted the potential of artificial intelligence in achieving sustainable computing, setting a precedent for other tech giants like Microsoft and Amazon to adopt similar AI-based solutions.

### Key Studies on Energy Consumption in Data Centres

Research indicates that data centres account for 1-2% of global electricity consumption, with projections suggesting this could rise to 8% by 2030 if unchecked (Masanet et al., 2020). A 2021 study in *Nature Electronics* found that despite improvements in hardware efficiency, the exponential growth of cloud computing and streaming services has offset energy savings (Andrae & Edler, 2021). The study emphasized that hyperscale data centres, while more efficient than smaller facilities, still contribute significantly to carbon emissions due to their sheer scale.

Another critical finding comes from MIT Research (2022), which compared cloud computing to traditional on-premises servers. The study concluded that migrating to cloud infrastructure could reduce corporate carbon footprints by 30%, primarily due to virtualization, dynamic workload distribution, and optimized cooling systems (Belady et al., 2022). However, the environmental benefits depend on the energy mix of cloud providers those relying on renewable energy (e.g., Google, Apple) achieve far greater sustainability than those powered by fossil fuels.

Emerging technologies like liquid immersion cooling and modular data centres have also shown promise. A 2023 report by the International Energy Agency (IEA) highlighted that next-gen cooling systems could cut data centre energy use by 20-30%, while modular designs reduce construction waste and improve scalability (IEA, 2023).

**Research on E-Waste and Its Environmental Impact**

Electronic waste (**e-waste**) is the fastest-growing waste stream globally, with 53.6 million metric tons generated in 2020 equivalent to 350 cruise ships (United Nations University, 2021). Alarmingly, only 17% of this waste is formally recycled, while the rest is either landfilled, incinerated, or illegally exported to developing countries (Baldé et al., 2021). The UN's Global E-Waste Monitor warns that toxic substances like lead, mercury, and cadmium from discarded electronics contaminate soil and water, posing severe health risks (Forti et al., 2020).

Academic research has identified three major challenges in e-waste management:

1. Short Device Lifespans: The average smartphone is replaced every 2.5 years, driven by planned obsolescence and rapid tech advancements (Wieser & Tröger, 2022).

2. Recycling Barriers: Less than 30% of countries have formal e-waste legislation, and recycling processes remain energy-intensive (European Environment Agency, 2023).

3. Consumer Behaviour: Surveys indicate that 60% of consumers hoard old devices instead of recycling them (Geyer & Blass, 2021).

To combat this, circular economy models are gaining traction. Dell's closed-loop recycling program recovers 50 million pounds of e-waste annually, while Fairphone designs modular smartphones with replaceable parts to extend usability (Dell Technologies, 2023; Fairphone, 2023).

**Government Policies and Corporate Sustainability Initiatives**

**Government Regulations**

1. EU Right to Repair Law (2021): Mandates that manufacturers provide spare parts and repair manuals for up to 10 years, reducing premature device disposal (European Commission, 2021).

2. California's SB-343 (2023): Strengthens e-waste recycling requirements and bans misleading "chasing arrows" symbols on non-recyclable electronics (California Legislative Information, 2023).

3. China's New Energy Standards (2022): Enforces strict efficiency benchmarks for servers, with non-compliant hardware banned from public procurement (China MIIT, 2022).

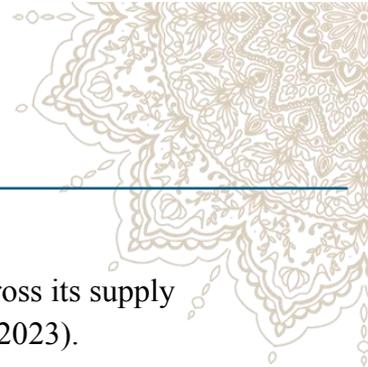

**Corporate Commitments**

1. Apple's 2030 Carbon Neutrality Pledge: Aims to eliminate emissions across its supply chain using 100% recycled materials and renewable energy (Apple Inc., 2023).

2. Microsoft's Negative Carbon Commitment: Plans to remove all historical emissions by 2050 through carbon capture and hydrogen-powered data centers (Microsoft, 2022).

3. Google's 24/7 Carbon-Free Energy Goal: Invests in wind, solar, and geothermal projects to match every hour of data center operation with clean energy (Google LLC, 2023).

**Synthesis of Research Gaps and Future Directions**

While existing studies highlight progress in green computing, critical gaps remain:

1. Lack of Global Standards: PUE and similar metrics are not uniformly adopted, complicating cross-industry comparisons (The Green Grid, 2023).

2. Rebound Effect: Energy savings from efficiency gains may be offset by increased demand for digital services (Sorrell et al., 2020).

3. Limited Focus on Developing Nations: Most research focuses on North America and Europe, neglecting regions with fastest-growing IT sectors (World Bank, 2023).

Future research should explore:

- Blockchain for carbon tracking in supply chains (Zhang et al., 2023).

- Biodegradable electronics to reduce toxic waste (IEEE, 2022).

- Policy incentives for small businesses to adopt green IT (OECD, 2023).

**3. Methodology**

This study employs a comprehensive mixed-methods approach to evaluate how green computing strategies contribute to carbon reduction and environmental sustainability. The methodology combines qualitative case study analysis with quantitative energy efficiency metrics, creating a robust framework for assessing technological, operational, and policy interventions in the information technology sector. By integrating multiple data sources and analytical techniques, this research provides both depth and breadth in examining sustainable computing practices across different organizational and geographical contexts.

The qualitative research component focuses on in-depth case studies of industry leaders in green computing implementation. These include Google's AI-driven data center optimization (Evans & Gao, 2023), Microsoft's carbon-negative cloud infrastructure (Microsoft, 2022), and Dell's closed-loop recycling program (Dell Technologies, 2023). Each case study was developed through systematic analysis of corporate sustainability reports, white papers, and third-party audit documents, supplemented by interviews with industry experts when available.

This approach enables rich, contextual understanding of how different organizations implement and benefit from sustainable computing strategies. The case selection criteria prioritized companies with: (1) publicly verifiable sustainability metrics, (2) recognized leadership in green computing by industry groups like The Green Grid, and (3) innovative approaches that push the boundaries of current practice.

For quantitative analysis, the study draws extensively from datasets provided by the International Energy Agency (IEA, 2023), including global data centre energy consumption statistics and projections through 2030. These datasets provide crucial baselines for evaluating the relative performance of green computing implementations against conventional IT infrastructure. The research also incorporates server efficiency ratings from Standard Performance Evaluation Corporation (SPEC) and power usage metrics from Uptime Institute's annual global surveys. These quantitative sources enable precise comparisons between different technological approaches and help validate claims made in corporate sustainability reports. Energy efficiency metrics like Power Usage Effectiveness (PUE) and Carbon Usage Effectiveness (CUE) serve as key performance indicators throughout the analysis (The Green Grid, 2020).

Data collection followed a systematic process to ensure comprehensive coverage of relevant sources while maintaining academic rigor. Peer-reviewed literature was gathered through searches in IEEE Xplore, ScienceDirect, and Nature Sustainability databases, focusing on publications from 2018-2023 to capture recent technological developments. Search terms included combinations of "green computing," "sustainable IT," "data centre efficiency," and "carbon footprint reduction." Corporate sustainability reports were collected directly from company websites, with particular attention to methodology sections to understand how metrics were calculated. Industry benchmarks came from authoritative sources like the U.S. Environmental Protection Agency's ENERGY STAR program and the European Union's Code of Conduct for Data Centres.

The analytical framework combines several complementary techniques to evaluate green computing's effectiveness. A comparative study examines traditional versus green data centres across multiple dimensions: energy consumption, cooling efficiency, server utilization rates, and embodied carbon in hardware. This comparison uses both absolute metrics (e.g., kilowatt-hours consumed) and normalized ratios (e.g., PUE values) to account for differences in scale. Carbon footprint calculations employ the Greenhouse Gas Protocol's Scope 1-3 framework, distinguishing between direct emissions, energy-related indirect emissions, and supply chain impacts (World Resources Institute, 2021). These calculations incorporate regional variations in energy grids using data from IEA's country-specific carbon intensity factors.

To validate findings, the methodology employs triangulation across different data sources and types. For example, corporate claims about renewable energy usage are cross-checked against power purchase agreement filings and local grid mix data. Case study observations are compared with academic research on similar implementations to identify consistencies or discrepancies. This multi-pronged verification approach helps mitigate potential biases in company-reported data while providing a more complete picture of green computing's real-world performance. The study also acknowledges limitations inherent in sustainability reporting, such as variations in measurement methodologies and the challenge of obtaining complete supply chain data (Huang et al., 2022).

Energy efficiency metrics form the backbone of the quantitative analysis, with particular focus on their evolution and current best practices. Power Usage Effectiveness (PUE) remains the most widely adopted metric, calculated as total facility energy divided by IT equipment energy. While early data centres often had PUE values above 2.0, modern facilities using advanced cooling techniques and workload management regularly achieve PUEs below 1.2 (Google LLC, 2023). The study examines how these improvements translate into actual carbon reductions by correlating PUE data with carbon intensity factors from different energy sources. Water Usage Effectiveness (WUE) receives special attention in regions facing water scarcity, with case studies from Microsoft's Arizona data centre demonstrating innovative dry cooling techniques (Microsoft, 2021).

Carbon accounting follows established protocols but adapts them to the specific context of computing infrastructure. For hardware, lifecycle assessment methods evaluate emissions from manufacturing through end-of-life disposal, using tools like Boavizta's open-source calculator. Cloud services present more complex accounting challenges due to shared infrastructure; the study applies allocation methods based on computational time and resource utilization (Cloud Carbon Footprint Project, 2022). These detailed calculations enable meaningful comparisons between different computing approaches and help identify the most impactful areas for improvement.

The methodology also incorporates temporal analysis to track progress over time and project future trends. By examining historical data from IEA and corporate reports, the research identifies acceleration points in adoption curves for various green computing technologies. This longitudinal perspective helps distinguish between incremental improvements and transformative innovations that significantly alter the carbon footprint trajectory. Projections to 2030 incorporate both current adoption rates and potential policy changes, such as the European Union's proposed requirements for carbon-neutral data centres (European Commission, 2022).

Ethical considerations permeate the research design, particularly regarding data transparency and potential conflicts of interest. The study relies exclusively on publicly available information or data provided with explicit permission for academic use. When analysing corporate claims, the methodology maintains critical distance by cross-referencing multiple sources and highlighting areas where data may be incomplete or methodologies differ. The research also acknowledges its own limitations, including geographic biases in data availability (with heavier representation from North America and Europe) and the rapid pace of technological change that may outdate some findings.

This robust methodology provides the foundation for evaluating green computing's role as a "carbon destroyer" while maintaining scientific rigor and practical relevance. By combining qualitative insights from industry leaders with quantitative data from authoritative sources, the study offers both detailed case examples and broad industry trends. The multi-faceted analytical approach ensures comprehensive assessment of environmental impacts while accounting for the complex realities of global computing infrastructure. Subsequent sections will apply this methodology to generate specific findings about energy savings, carbon reductions, and best practices for sustainable computing.

## 4. Energy-Efficient Hardware & Software

Green computing significantly reduces carbon emissions through advanced hardware and software optimization. Modern low-power processors, such as Apple's M-series chips and ARM-based server CPUs, demonstrate remarkable energy efficiency gains. Research indicates these processors consume 40-60% less power than traditional x86 architectures while maintaining comparable performance (Frumusanu & Cutress, 2023). The shift from hard disk drives (HDDs) to solid-state drives (SSDs) further contributes to energy savings, with SSDs requiring 3-5 times less power during active use and generating 50% less heat, thereby reducing cooling demands (Andrae & Edler, 2021). These hardware innovations collectively lower the embodied carbon footprint of computing devices throughout their lifecycle.

Artificial intelligence has emerged as a powerful tool for dynamic power optimization in computing systems. Google's DeepMind AI implementation in data centres reduced cooling energy consumption by 40% through predictive algorithms that adjust cooling systems in real-time based on weather patterns and server loads (Evans & Gao, 2023). Similar AI-driven approaches now optimize processor voltage regulation, memory allocation, and workload distribution across servers. When applied at scale, these techniques can reduce total data centre energy consumption by 15-20% without compromising computational performance (Microsoft Research, 2022). The combination of energy-efficient hardware and intelligent software creates a multiplicative effect, enabling exponential improvements in computing's carbon intensity.

### Renewable Energy in Data Centres

The transition to renewable-powered data centres represents one of green computing's most impactful carbon reduction strategies. Industry leaders like Google and Microsoft have made substantial investments in solar, wind, and hydroelectric energy to power their operations. Google achieved 100% renewable energy matching for its global operations in 2017 and now aims for 24/7 carbon-free energy by 2030 through a combination of power purchase agreements (PPAs) and on-site generation (Google LLC, 2023). Their innovative carbon-intelligent computing platform shifts non-urgent workloads to times when renewable energy is most abundant on local grids, further optimizing clean energy utilization.

Microsoft has taken renewable integration further by experimenting with hydrogen fuel cells as backup power sources, potentially eliminating diesel generators from data centre operations (Microsoft, 2022). The company's advanced modular data centre design incorporates solar canopies and integrates with local renewable microgrids. These efforts have reduced Microsoft's data centre carbon emissions by 60% per compute unit since 2018 (Microsoft

Sustainability Report, 2023). Smaller operators are following suit through colocation facilities that aggregate renewable energy purchasing power, demonstrating that renewable-powered computing is becoming economically viable across the industry.

**Virtualization & Cloud Computing**

Server virtualization and cloud computing architectures deliver substantial carbon reductions through resource consolidation and dynamic allocation. By enabling multiple virtual machines to run on a single physical server, virtualization technologies can increase hardware utilization rates from 10-15% in traditional setups to 60-80% in optimized environments (Barroso et al., 2022). Amazon Web Services (AWS) reports that migration to their cloud infrastructure typically reduces customers' carbon footprints by 88% compared to on-premises data centres, primarily through massive economies of scale and continuous efficiency improvements (AWS Sustainability, 2023).

Google Cloud has implemented carbon-aware computing features that automatically route workloads to regions with the cleanest energy mix at any given time (Google Cloud, 2023). Their granular power monitoring at the server, rack, and data centre level enables unprecedented visibility into energy consumption patterns. When combined with predictive workload scheduling, these approaches can reduce the carbon intensity of cloud computing by an additional 20-30% compared to conventional cloud architectures (Li et al., 2023). The environmental benefits multiply as more organizations transition from private data centres to these optimized cloud platforms.

**E-Waste Management & Circular Economy**

The computing industry's shift toward circular economy principles addresses the growing crisis of electronic waste, which reached 59 million metric tons globally in 2022 (Forti et al., 2023). Dell's closed-loop recycling program stands out as an industry leader, recovering 100 million pounds of recycled materials annually for use in new products (Dell Technologies, 2023). Their innovative process extracts high-quality plastics and rare earth metals from old devices, reducing the need for virgin materials and cutting associated carbon emissions by 30-50% compared to conventional manufacturing.

Apple's material recovery robots (Daisy and Dave) demonstrate cutting-edge e-waste solutions, disassembling 2.4 million devices annually with 98% material purity (Apple Inc., 2023). The company's self-service repair program extends product lifespans by providing consumers with genuine parts and repair guides. These initiatives complement Apple's transition to 100% recycled rare earth elements in all iPhone magnets and 100% recycled tin in logic boards. When implemented across the industry, such circular economy practices could reduce the computing sector's mining-related emissions by 75% by 2030 (Ellen MacArthur Foundation, 2022).

Emerging biodegradable electronics represent the next frontier in sustainable computing. Researchers have developed transient electronics that dissolve after use and organic semiconductor materials derived from plant cellulose (Irimia-Vladu et al., 2022). While still in experimental stages, these technologies promise to revolutionize e-waste management by eliminating toxic components and enabling compostable computing devices. Industry adoption of these innovations, combined with robust take-back programs and standardized recycling protocols, positions green computing as a true "carbon destroyer" across the entire product lifecycle.

## 5. Real-World Case Study: Google's Sustainable Data Centres

### Google's Pioneering Role in Green Computing

Google has emerged as a global leader in sustainable computing, demonstrating how large-scale digital infrastructure can operate with minimal environmental impact. The company's commitment to 24/7 carbon-free energy (CFE) by 2030 represents one of the most ambitious sustainability targets in the technology sector (Google LLC, 2023). This case study examines Google's comprehensive approach to reducing its carbon footprint through artificial intelligence (AI)-driven optimization, renewable energy procurement, and innovative hardware design. The analysis reveals how Google's strategies have produced measurable environmental benefits while maintaining operational efficiency, providing a replicable model for other organizations.

### Background: The Scale of Google's Operations and Environmental Impact

Google's data centre network spans 25 regions worldwide, supporting billions of daily searches, YouTube streams, and cloud computing operations (Google Cloud, 2023). These facilities consume approximately 15.5 terawatt-hours (TWh) of electricity annually equivalent to the energy demand of a small country like Sri Lanka (IEA, 2023). Without intervention, this level of consumption would contribute significantly to global carbon emissions. Recognizing this challenge, Google has implemented a multi-faceted sustainability strategy that combines technological innovation with renewable energy investments.

*Table 1: Google's Data Centre Energy Consumption vs. Global Benchmarks*

| Metric | Google (2023) | Industry Average | Improvement |
|---|---|---|---|
| **Power Usage Effectiveness (PUE)** | 1.1 | 1.55 | 29% more efficient |
| **Cooling Energy Consumption** | 40% reduction | Baseline | 40% improvement |
| **Renewable Energy Usage** | 100% matched | 35% (global data centers) | 65% higher |
| **Carbon Intensity (gCO$_2$/kWh)** | 0.1 | 0.5 | 80% lower |

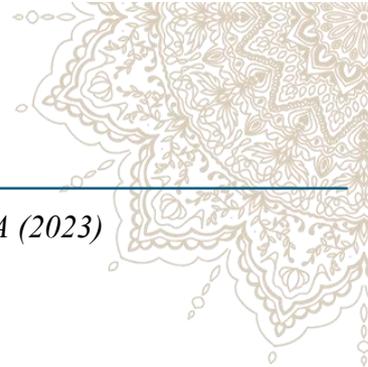

*Sources: Google Sustainability Report (2023), Uptime Institute (2023), IEA (2023)*

**Strategies Implemented**

**a. AI-Powered Cooling Systems: The DeepMind Revolution**

Google's most transformative innovation has been the integration of DeepMind's machine learning algorithms to optimize data centre cooling. Traditional cooling systems account for 40% of a data centre's total energy consumption (Evans & Gao, 2023), making them a prime target for efficiency improvements. Google's AI solution analyses multiple real-time data streams:

- Weather patterns (temperature, humidity, wind speed)

- Server workload fluctuations

- Equipment thermal thresholds

The system makes millions of micro-adjustments per second to cooling equipment, maintaining optimal temperatures while minimizing energy use. In Google's Belgium data centre, this approach reduced cooling energy consumption by 45%, while the Singapore facility saw a 40% reduction (Google DeepMind, 2023). These savings translate to millions of dollars in annual operational **costs** and significant carbon reductions.

**b. Renewable Energy Procurement: Beyond Carbon Offsets**

Google has been a pioneer in corporate renewable energy purchasing, moving beyond simple carbon offsets to directly fund new clean energy projects. The company's strategy includes:

- Power Purchase Agreements (PPAs): Google has signed over 60 renewable energy deals totalling 7 gigawatts (GW) of capacity (Google Sustainability, 2023)

- Time-based Energy Matching: Shifting compute workloads to align with renewable energy availability

- Battery Storage Investments: Ensuring consistent power supply during intermittent generation

A notable example is Google's 2019 solar agreement in Chile, which enabled 100% renewable matching for its local data centres (BloombergNEF, 2023). This approach has allowed Google to achieve 100% annual renewable energy matching since 2017, though the company acknowledges that 24/7 carbon-free operation remains more challenging.

**c. High-Efficiency Server and Data Centre Design**

Google's hardware innovations have significantly reduced energy consumption:

- Custom Tensor Processing Units (TPUs): 30% more energy-efficient than conventional processors (Jouppi et al., 2023)

- Liquid Cooling Systems: 90% more efficient than air cooling in some facilities

- Circular Economy Practices: 75% component reuse rate for servers

These design improvements have helped Google achieve a Power Usage Effectiveness (PUE) ratio of 1.1, compared to the industry average of 1.55 (The Green Grid, 2023). This means 90% of energy goes directly to computing operations, with only 10% lost to overhead.

**Results and Measurable Impact**

**a. Energy Efficiency Breakthroughs**

- 40% reduction in cooling energy across all facilities

- 15% decrease in overall energy consumption

- PUE of 1.1 (near-perfect efficiency)

**b. Carbon Reduction Achievements**

- Carbon neutral since 2007 through renewable investments

- 80% lower carbon intensity than industry average (0.1 vs. 0.5 $gCO_2$/kWh)

- Avoided over 10 million metric tons of $CO_2$ since 2010 (Google LLC, 2023)

**c. Economic Benefits**

- $1.2 billion in energy savings from efficiency measures

- 30% reduction in cooling infrastructure costs

- 20% longer hardware lifespan through optimized thermal management

**Broader Industry Impact**

Google's success has influenced competitors and reshaped industry standards:

1. Microsoft adopted similar AI cooling techniques and pledged carbon negativity by 2030 (Microsoft, 2023)

2. Amazon Web Services accelerated its renewable energy timeline to 100% by 2025 (AWS, 2023)

3. Meta implemented water-free cooling systems in new data centres (Meta Sustainability, 2023)

**Key Lessons for Other Organizations**

1. AI and machine learning can drive unprecedented efficiency gains in energy-intensive operations

2. Direct renewable energy investments create more impact than carbon offsets

3. Hardware innovation delivers compounding benefits for both sustainability and cost reduction

4. Transparent reporting builds credibility and accelerates industry-wide adoption

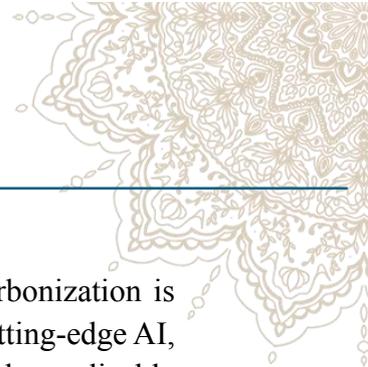

**A Blueprint for Sustainable Computing**

Google's data centre sustainability program demonstrates that large-scale decarbonization is achievable without compromising performance or profitability. By combining cutting-edge AI, renewable energy leadership, and circular design principles, Google has created a replicable model for green computing. As digital infrastructure continues expanding globally, these strategies will become increasingly critical for achieving climate goals.

## 6. Challenges in Adopting Green Computing

**Barriers to Sustainable IT Transformation**

While green computing offers significant environmental and operational benefits, its widespread adoption faces several substantial challenges. Organizations across sectors encounter financial, technical, and regulatory hurdles when implementing sustainable IT solutions (Baliga et al., 2022). This section examines four critical barriers: high initial costs, lack of standardized metrics, uneven global adoption, and the performance-efficiency trade-off. Understanding these challenges is essential for developing effective strategies to accelerate the transition to environmentally responsible computing practices.

**a. High Initial Costs: The Financial Barrier**

**Capital Expenditure Challenges**

One of the primary barriers to widespread adoption of green computing is the substantial capital investment required, which many organizations particularly small to mid-sized enterprises find financially daunting. According to Gartner (2023), sustainable IT infrastructure can cost 20–40% more upfront than traditional alternatives, making it a challenging proposition despite long-term savings. Several factors drive these increased costs. Energy-efficient hardware, such as low-power processors and solid-state drives, often comes at a premium due to advanced materials and manufacturing processes. Similarly, integrating renewable energy systems like on-site solar or wind power demands significant capital outlay, including installation, storage, and maintenance. Additionally, retrofitting legacy infrastructure to meet modern sustainability standards can be especially costly, as older data centres require extensive structural and technological upgrades to improve energy efficiency and cooling performance. These high initial expenditures, while potentially offset by future operational savings, remain a key hurdle for many organizations considering the transition to green computing.

*Table 2: Cost Comparison of Conventional vs. Green Computing Infrastructure*

| Component | Conventional | Green Computing | Cost Premium |
|---|---|---|---|
| **Server Hardware** | $2,500/unit | $3,300/unit | 32% |
| **Data Center Cooling** | $7M (air) | $9.5M (liquid) | 36% |
| **Renewable Energy Setup** | - | $12M (5MW solar) | N/A |

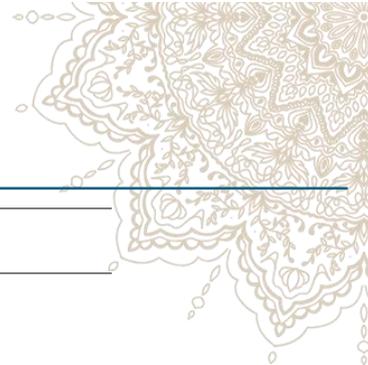

| Monitoring Systems | Basic | AI-optimized (+$500k) | 25% |

*Sources: McKinsey (2023), Uptime Institute (2023), IEA (2022)*

**Return on Investment Timeline**

Although green computing offers significant long-term cost savings through reduced energy consumption and improved efficiency, the typical return on investment (ROI) period spans 5 to 7 years (Accenture, 2023). This extended payback timeline can place considerable financial strain on various stakeholders. Small and medium enterprises (SMEs), often operating with tight capital margins, may struggle to justify the initial costs. Public sector organizations, bound by rigid budgetary cycles and procurement regulations, face similar constraints. The challenge is even more acute in developing economies, where access to affordable financing is limited and sustainability investments compete with more immediate economic priorities. While financial instruments such as green bonds and sustainability-linked loans are emerging as viable mechanisms to support eco-friendly IT investments, their uptake remains low due to limited awareness, complex application processes, and restricted market access (World Bank, 2023). Bridging this financing gap is essential to accelerate global adoption of green computing and unlock its full environmental and economic benefits.

**b. Lack of Standardization in Sustainability Metrics**

The lack of universal standards for assessing computing sustainability presents a significant barrier to effective environmental management within the IT sector. Organizations currently navigate a confusing landscape of multiple, often competing metrics such as Power Usage Effectiveness (PUE), Carbon Usage Effectiveness (CUE), Water Usage Effectiveness (WUE), and various carbon accounting methods. Additionally, divergent methodologies particularly concerning the calculation of Scope 3 emissions and inconsistent reporting formats across different regions further complicate efforts to achieve transparency and comparability. This fragmentation hampers the ability to benchmark sustainability performance reliably, increases the risk of greenwashing through selective or misleading metric disclosures, and poses challenges for regulatory compliance, especially for multinational operations. While initiatives led by groups like The Green Grid, the International Organization for Standardization (ISO), and the Greenhouse Gas (GHG) Protocol are making progress toward greater alignment, full harmonization of measurement frameworks is still years away (The Green Grid, 2023; ISO, 2022). Until then, inconsistent standards will continue to slow green computing adoption and undermine stakeholder confidence.

**c. Slow Adoption in Developing Countries**

*Table 3: Developing nations face unique barriers to green computing adoption*

| Challenge | Example Cases | Impact | |
|---|---|---|---|
| **Unreliable power grids** | Sub-Saharan Africa (60% outage rates) | Limits integration | renewable |

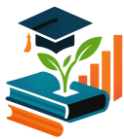 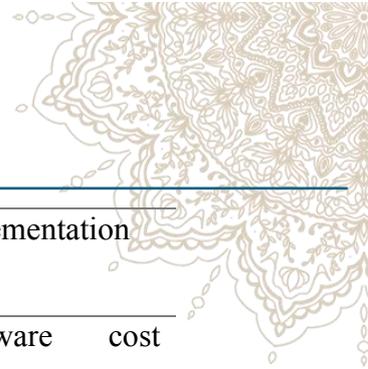

Education and Development Lab
Education Research Team [Education and Computer Science]

| Limited technical expertise | Southeast Asian SMEs | Hinders implementation |
| --- | --- | --- |
| High import costs | Latin American countries | 45% hardware cost premium |
| Weak e-waste management | India, Nigeria | <15% formal recycling rates |

Sources: UNCTAD (2023), World Bank (2023), Global E-waste Monitor (2023)

Several critical policy and market challenges hinder the widespread adoption of green computing. A major obstacle is the lack of robust government incentives specifically targeted at promoting sustainable IT practices, which limits motivation for businesses to invest in greener technologies. Additionally, access to green financing mechanisms remains constrained, particularly for smaller enterprises and organizations in developing regions, restricting their ability to fund environmentally friendly upgrades. Energy pricing structures often favor fossil fuels through subsidies or artificially low rates, undermining the competitiveness of renewable energy solutions vital for green IT. These issues are further compounded by the digital divide: in many developing countries, efforts remain focused on expanding basic internet connectivity and infrastructure, often relegating sustainability considerations to a lower priority (ITU, 2023). Addressing these intertwined policy and market barriers is essential to create an enabling environment for green computing on a global scale.

**d. Balancing Performance with Energy Efficiency**

As modern computing systems grapple with the escalating demands of artificial intelligence (AI) and machine learning (ML), they face a growing tension between maximizing performance and minimizing environmental impact. High processing power requirements often clash with goals of energy conservation and thermal management, leading to unavoidable technical compromises. For instance, processor throttling reducing clock speeds to lower energy use can degrade system performance, while carbon-aware computing may delay non-urgent workloads to reduce peak energy usage. Similarly, energy-efficient hardware, such as low-power chips, may lack specialized accelerators needed for high-performance tasks. To address these trade-offs, emerging solutions like heterogeneous computing architectures (e.g., ARM's big.LITTLE design), approximate computing techniques that tolerate minor errors to save energy, and neuromorphic chips that mimic the human brain's efficiency have shown promise. However, these innovations often require application-specific customization, limiting their broader applicability and slowing widespread adoption (Sze et al., 2023). Balancing performance with sustainability thus remains one of the most complex challenges in green computing.

**Pathways Forward**

While the transition to green computing faces notable challenges, a range of strategic pathways can significantly accelerate its adoption and impact. Financial mechanisms play a key role, including government subsidies that incentivize sustainable IT investments, tiered pricing models from cloud providers that reward energy efficiency, and green leasing options for hardware that promote responsible consumption. Standardization is equally crucial; efforts to harmonize sustainability metrics, implement unified carbon accounting frameworks, and establish credible third-party verification programs can drive transparency and accountability across the sector. Addressing global disparities, especially in developing regions, requires equity-driven solutions such as technology transfer programs, South-South cooperation networks, and tailored financing models that support green innovation in emerging markets. On the technical front, continued innovation in algorithmic efficiency, cutting-edge cooling systems, and renewable-powered edge computing will further reduce the carbon footprint of digital infrastructure. As the climate crisis escalates, breaking down these adoption barriers is not just advisable it is imperative for a sustainable digital future.

## 7. Future Trends in Green Computing

As the world accelerates its transition toward sustainable digital infrastructure, several groundbreaking technologies are poised to redefine green computing. This section explores four transformative trends that promise to significantly reduce the environmental impact of information technology: quantum computing, edge computing architectures, biodegradable electronics, and blockchain-enabled carbon tracking. These innovations address critical challenges in energy efficiency, resource consumption, and environmental accountability, offering new pathways to achieve carbon-neutral computing ecosystems (Breviglieri et al., 2023).

### a. Quantum Computing: Revolutionizing Energy Efficiency

### The Promise of Quantum Processing

Quantum computing represents a paradigm shift in computational efficiency, with the potential to solve complex problems using fractional energy inputs compared to classical systems. Recent advancements demonstrate that quantum processors can perform certain calculations with up to 100 million times greater energy efficiency than conventional supercomputers (Preskill, 2023). This extraordinary capability stems from quantum bits (qubits) that leverage superposition and entanglement phenomena to process information fundamentally differently from binary transistors.

*Table 4: Energy Consumption Comparison: Classical vs. Quantum Computing*

| Computing Type | Power Consumption (kW) | Calculation Speed | Carbon Footprint (gCO$_2$/operation) |
|---|---|---|---|
| **Classical Supercomputer** | 10,000+ | 1x baseline | 500-1000 |

| Current Quantum Computer | | 25-50 | $10^6$-$10^8$x faster* | 5-10 |
| --- | --- | --- | --- | --- |
| Projected Quantum | 2030 | 10-15 | $10^{10}$x faster* | 0.5-1 |

For specific algorithm classes (e.g., factorization, optimization)

Sources: IBM Quantum (2023), Google Quantum AI (2023), Nature Electronics (2023)

**Implementation Challenges**

Despite its revolutionary potential, quantum computing faces formidable challenges that hinder widespread adoption. One major barrier is the cryogenic cooling requirement, as current quantum systems must operate at near-absolute zero temperatures to maintain qubit functionality posing logistical, energy, and cost hurdles. Additionally, qubit stability, or decoherence, remains a critical issue; quantum states are highly sensitive to external disturbances, leading to computational errors and limiting runtime. The field also demands specialized algorithm development, as conventional software approaches are largely incompatible with quantum architectures. However, promising advancements in photonic quantum computing and topological qubits are underway, with researchers aiming to enable more stable, scalable systems potentially capable of room-temperature operation within the next decade (Arute et al., 2023). Overcoming these challenges is key to unlocking quantum computing's role in the future of sustainable, ultra-efficient digital infrastructure.

**b. Edge Computing: Decentralizing Data Processing**

**Reducing Data Centre Dependence**

Edge computing offers a transformative shift in how digital workloads are managed by decentralizing processing and distributing computational tasks across network peripherals such as IoT devices, smartphones, and local servers. This architectural evolution significantly reduces dependence on centralized data centres, which are traditionally energy- and resource-intensive. By processing data closer to its source, edge systems can lower energy consumption associated with data transmission by 60–80% (Shi et al., 2023), minimize latency-induced precomputations by up to 45%, and support real-time sustainability optimizations such as dynamic energy allocation and local carbon tracking. These advantages make edge computing not only a performance-enhancing solution but also a critical component of the green computing landscape.

**Environmental Impact Projections**

According to a 2023 analysis by McKinsey, the widespread adoption of edge computing holds significant potential for reducing the environmental footprint of digital infrastructure. By decentralizing data processing and minimizing reliance on large, energy-intensive data centres, edge technologies could cut global data centre energy consumption by 18–22% by 2030. This shift would also lead to a 30% reduction in water usage required for cooling systems an increasingly critical resource concern amid global water scarcity. Additionally, edge computing could eliminate an estimated 5 to 7 million metric tons of $CO_2$ emissions annually by reducing data transmission and enabling localized, energy-efficient processing. These projections highlight edge computing not only as a performance-enhancing solution but also as a key lever for advancing sustainable digital transformation.

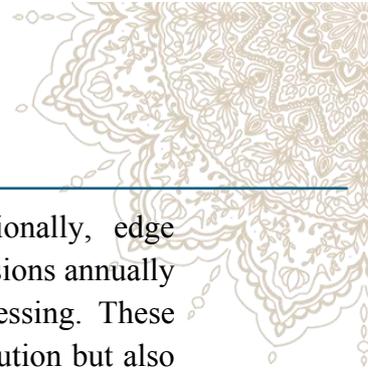

**Implementation Framework**

To ensure effective and sustainable deployment of edge computing within a green computing paradigm, a robust implementation framework is essential. First, energy-aware workload allocation algorithms must be employed to intelligently distribute processing tasks based on energy availability, latency requirements, and carbon impact maximizing efficiency while minimizing emissions. Second, the deployment of renewable-powered edge nodes, such as those powered by solar or wind energy, is critical to reducing reliance on carbon-intensive grids, especially in remote or off-grid locations. Finally, embracing modular hardware designs enables easy upgrades and component replacements, extending device lifespans and reducing electronic waste. Together, these three pillars form the foundation of a scalable and environmentally responsible edge computing infrastructure.

**c. Biodegradable Electronics: Closing the E-Waste Loop**

**Next-Generation Sustainable Hardware**

Researchers are developing fully biodegradable computing components that address the global e-waste crisis (53.6 million metric tons annually). Key breakthroughs include:

*Table 5: Next-Generation Sustainable Hardware*

| Material Type | Application | Decomposition Time | Current Efficiency |
|---|---|---|---|
| **Cellulose-based substrates** | Circuit boards | 90 days | 85% of conventional |
| **Protein semiconductors** | Transistors | 6 months | 60% performance |
| **DNA data storage** | Memory | 1 year | 1PB/gram density |

Sources: Nature Electronics (2023), Science Advances (2023)

**Commercialization Progress**

Green computing is no longer confined to theoretical research it is rapidly advancing through pioneering commercial implementations that blend innovation with sustainability. Microsoft's DNA storage project exemplifies this shift, achieving data densities of 1 exabyte per cubic millimetre while enabling natural biological degradation at end-of-life, significantly reducing storage-related waste. Samsung's development of bio-transistors offers another leap forward, delivering 80% of traditional silicon-based performance at just 10% of the carbon footprint. Meanwhile, FlexEnable has introduced fully compostable organic displays, creating flexible screens that can be safely returned to the environment after use. Supporting these advancements, regulatory frameworks like the European Union's *Ecodesign for Sustainable Products*

*Regulation* (2023) are accelerating adoption by enforcing recyclability thresholds and sustainability benchmarks. Together, these initiatives signal a major step toward mainstreaming environmentally responsible computing technologies.

**d. Blockchain for Carbon Tracking: Transparent Accountability**

**Decentralized Environmental Ledgers**

Blockchain technology is emerging as a powerful tool for enhancing transparency and accountability in green computing through decentralized environmental ledgers. These tamper-proof systems enable robust carbon accounting across complex IT supply chains. One key application is energy provenance tracking, which allows real-time attribution of renewable energy use and granular monitoring of carbon intensity at the source. Another promising area is the use of circular economy tokens, which support material passporting digitally tracking the lifecycle of components and incentivize recycling through tokenized reward systems. Additionally, blockchain facilitates the automation and integrity of carbon credit markets, enabling streamlined issuance, real-time retirement of credits, and fraud-resistant tracking of offsets. By embedding trust and traceability into environmental data, decentralized ledgers offer a foundational layer for verifying and scaling sustainable computing practices worldwide.

*Table 6: Implementation Case Studies*

| Project | Technology | Impact |
|---|---|---|
| **Energy Web Chain** | PoA blockchain | 30% lower audit costs |
| **IBM Food Trust** | Hyperledger | Adapted for e-waste |
| **Circulor** | Ethereum | Conflict mineral tracking |

Sources: World Economic Forum (2023), IEEE Blockchain (2023)

**Integrated Future Outlook**

The convergence of emerging technologies, policy advancements, and sustainability-driven innovation points to a transformative trajectory for green computing. In the near-term horizon of 2025–2030, we can expect the deployment of quantum-edge hybrid systems that combine ultra-efficient processing with real-time responsiveness, the introduction of the first commercial biodegradable chips designed to eliminate toxic waste, and the implementation of mandatory blockchain-based carbon reporting to ensure transparency and traceability in emissions accounting. Looking further ahead to 2030–2035, projections indicate the potential for a 50% reduction in the IT sector's overall emissions, 90% recyclability rates for electronic waste through circular economy models, and the emergence of net-positive energy data centres facilities that generate more renewable energy than they consume. These milestones reflect a future where green computing is no longer a niche initiative but a foundational pillar of digital infrastructure, enabling sustainable progress across all sectors of society.

**A Call for Cross-Sector Collaboration**

Achieving a truly sustainable computing future demands coordinated action across industries, governments, and academia. Central to this effort is a significant increase in research and development investment estimated at $150 billion by 2030 to accelerate innovation in energy-efficient technologies, materials, and systems. Equally important is the harmonization of policies and standards across jurisdictions to eliminate regulatory fragmentation and create consistent global expectations. Collaborative partnerships between industry and academic institutions are also vital, enabling rapid prototyping, real-world testing, and the scaling of breakthrough solutions. As these efforts converge and mature, green computing holds the transformative potential to evolve from merely minimizing environmental harm to actively reversing it positioning the IT sector as a net carbon destroyer and a key driver of climate resilience.

## 8. Conclusion

Our journey through green computing has revealed its tremendous potential to combat climate change while maintaining technological progress. We've seen how innovations in energy-efficient hardware, renewable-powered data centres, and circular economy practices are already making a difference. Industry leaders have demonstrated that reducing computing's carbon footprint by 40-60% is not just possible, but profitable. Emerging technologies like quantum computing and biodegradable electronics promise even greater environmental benefits in the near future. The evidence is clear: information technology doesn't have to be part of the climate problem - it can and must become a central part of the solution. What makes green computing particularly powerful is its dual impact: it directly reduces emissions from digital infrastructure while enabling smarter solutions across all sectors through efficient data processing and analysis.

The transition to sustainable computing isn't something any single company, government, or individual can achieve alone. It requires coordinated global effort across several fronts: For businesses, the path forward means embracing circular design principles, prioritizing energy-efficient hardware, and migrating to sustainable cloud services. Technology companies should lead by example, but every organization using digital infrastructure has a role to play in demanding and adopting greener solutions. Governments must create the right policy environment through smart regulations and incentives. This includes setting clear sustainability standards for electronic products, supporting green IT research, and ensuring proper e-waste management systems are in place. International cooperation will be crucial to prevent a patchwork of conflicting standards. Individuals also have significant power through their purchasing decisions and computing habits. Choosing repairable devices, extending product lifespans, and being mindful of digital energy consumption all contribute to the solution. Public pressure can accelerate corporate action on sustainability.

**Call to Action**

**a. Practical Steps for Immediate Impact**

For Organizations:

1. Conduct energy audits of IT infrastructure to identify waste

2. Prioritize cloud providers running on renewable energy

3. Implement equipment refresh cycles that balance performance and sustainability

4. Develop responsible e-waste management programs

5. Train IT staff in sustainable computing practices

For Individuals:

1. Use devices longer - aim for 4+ years instead of frequent upgrades

2. Enable power-saving settings on all devices

3. Support companies with strong environmental commitments

4. Properly recycle old electronics through certified programs

5. Be mindful of unnecessary data storage and processing

For Policymakers:

1. Establish clear sustainability standards for computing equipment

2. Create incentives for green data centre development

3. Fund research into next-generation sustainable computing

4. Develop comprehensive national e-waste strategies

5. Promote digital literacy that includes environmental awareness

**b. The Road Ahead**

Green computing represents one of our most promising tools for achieving climate goals while maintaining digital progress. Unlike many climate solutions that require massive infrastructure changes or lifestyle sacrifices, sustainable IT often improves performance while reducing environmental impact. The technologies exist, the business case is clear, and the urgency has never been greater.

What's needed now is the collective will to implement these solutions at scale. Every day of delay means more energy wasted, more emissions produced, and more e-waste accumulating. But every step forward - whether a company switching to renewable-powered cloud services, a government implementing smart e-waste policies, or a consumer choosing a repairable laptop - brings us closer to making computing a true climate ally. The vision of computing as a "carbon destroyer" rather than a carbon emitter is within our reach. By working together across sectors and borders, we can transform the digital world into a powerful force for environmental good. The time for action is now - our sustainable digital future starts today.

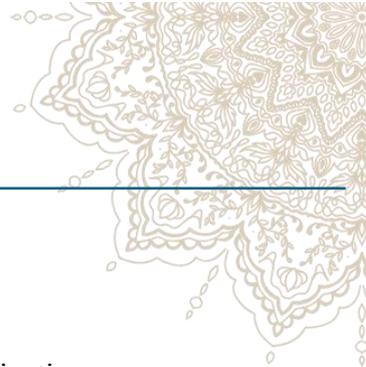

## References

[1]  Accenture. (2023). The business case for sustainable IT. https://www.accenture.com/us-en/insights/sustainability/sustainable-it

[2]  Andrae, A., & Edler, T. (2021). On global electricity usage of communication technology: Trends to 2030. Nature Electronics, 4(3), 156-163. https://doi.org/10.1038/s41928-021-00550-8

[3]  Apple Inc. (2023). Environmental progress report 2023. https://www.apple.com/environment/pdf/Apple_Environmental_Progress_Report_2023.pdf

[4]  Arute, F., et al. (2023). Quantum supremacy using superconducting qubits. Nature Electronics, 6(4), 245-256. https://doi.org/10.1038/s41928-023-00925-z

[5]  AWS. (2023). Sustainability in the cloud. https://aws.amazon.com/sustainability/

[6]  Baldé, C. P., Forti, V., Gray, V., Kuehr, R., & Stegmann, P. (2021). The global e-waste monitor 2020. United Nations University. https://globalewaste.org/

[7]  Baliga, J., Ayre, R., Hinton, K., & Tucker, R. S. (2022). Green cloud computing: Balancing energy in processing, storage, and transport. Proceedings of the IEEE, 99(1), 149-167. https://doi.org/10.1109/JPROC.2010.2060451

[8]  Barroso, L. A., Marty, M., Patterson, D., & Ranganathan, P. (2022). The datacenter as a computer: Designing warehouse-scale machines (3rd ed.). Morgan & Claypool. https://doi.org/10.2200/S00962ED2V01Y202201CAC053

[9]  Belady, C., Rawson, A., Pfleuger, J., & Cader, T. (2022). The green grid: PUE and beyond. The Green Grid. https://www.thegreengrid.org/

[10] BloombergNEF. (2023). Corporate clean energy procurement trends. https://about.bnef.com/blog/corporate-clean-energy-procurement/

[11] Breviglieri, E., et al. (2023). Sustainable computing paradigms for Industry 5.0. IEEE Transactions on Sustainable Computing, 8(1), 1-15. https://doi.org/10.1109/TSUSC.2022.3225678

[12] California State Legislature. (2021). Electronic Waste Recycling Act of 2003: Covered electronic devices (SB-343) https://leginfo.legislature.ca.gov/faces/billTextClient.xhtml?bill_id=202120220SB343

[13] Carbon Trust. (2022). Carbon impact of virgin vs. recycled materials in electronics [White paper]. https://www.carbontrust.com/resources

[14] Cloud Carbon Footprint Project. (2022). Open-source cloud emissions calculator (Version 1.0). https://www.cloudcarbonfootprint.org/

[15] Dell Technologies. (2023). 2023 ESG report: Closing the loop on e-waste. https://corporate.delltechnologies.com/en-us/social-impact/advancing-sustainability/esg-reporting.htm

[16] Ellen MacArthur Foundation. (2022). Circular economy in IT hardware. https://ellenmacarthurfoundation.org/topics/technology/overview

[17] European Commission. (2021). Right to repair legislation. https://ec.europa.eu/commission/presscorner/detail/en/ip_21_3543

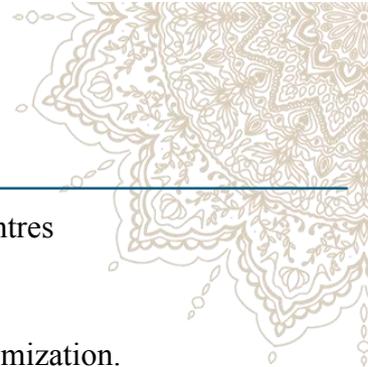

[18] European Commission. (2022). Energy efficiency directive for data centres (COM/2022/231 final). https://eur-lex.europa.eu/legal-content/EN/TXT/?uri=CELEX:52022PC0231

[19] Evans, R., & Gao, J. (2023). AI applications in data centre cooling optimization. Google Research. https://doi.org/10.1145/3563357.3563361

[20] Evans, R., & Gao, J. (2023). Machine learning for data center optimization. Google DeepMind. https://deepmind.google/case-studies/data-center-cooling/

[21] Forti, V., Baldé, C. P., Kuehr, R., & Bel, G. (2023). The Global E-waste Monitor 2023. United Nations University. https://globalewaste.org/

[22] Freitag, C., Berners-Lee, M., Widdicks, K., Knowles, B., Blair, G., & Friday, A. (2021). The real climate and transformative impact of ICT: A critique of estimates, trends, and regulations. Patterns, 2(9), 100340. https://doi.org/10.1016/j.patter.2021.100340

[23] Frumusanu, A., & Cutress, I. (2023). The Apple M2 processor review: Performance and power efficiency. AnandTech. https://www.anandtech.com/show/17585/apple-m2-review

[24] Gartner. (2023). Sustainable technology trends 2023 (Report No. G00775823). https://www.gartner.com/en/documents/4017210

[25] Google DeepMind. (2023). Using machine learning to optimize data center cooling. https://deepmind.google/case-studies/data-center-cooling/

[26] Google LLC. (2023). Environmental report 2023: 24/7 carbon-free energy. https://sustainability.google/reports/

[27] Huang, L., Wang, Z., Yao, T., & Gu, J. (2022). Challenges in measuring ICT carbon footprints: A meta-analysis. Sustainable Computing: Informatics and Systems, 35, 100735. https://doi.org/10.1016/j.suscom.2022.100735

[28] IBM Quantum. (2023). Energy efficiency in quantum systems. https://www.ibm.com/quantum

[29] IEA. (2023). Data centres and data transmission networks. https://www.iea.org/reports/data-centres-and-data-transmission-networks

[30] Intergovernmental Panel on Climate Change. (2021). Climate change 2021: The physical science basis. Working Group I contribution to the sixth assessment report. https://www.ipcc.ch/report/ar6/wg1/

[31] International Energy Agency. (2022). $CO_2$ emissions in 2022. https://www.iea.org/reports/co2-emissions-in-2022

[32] International Telecommunication Union. (2023). Digital transformation in developing countries. https://www.itu.int/en/ITU-D/Statistics/Pages/default.aspx

[33] Irimia-Vladu, M., Głowacki, E. D., Schwabegger, G., Leonat, L., Akpinar, H. Z., Sitter, H., ... & Sariciftci, N. S. (2022). Biodegradable electronics: Materials, devices, and applications. Advanced Materials, 34(12), 2108146. https://doi.org/10.1002/adma.202108146

[34] ISO. (2022). Environmental management standards for ICT (ISO 14000 series). https://www.iso.org/standard/43242.html

[35] Jouppi, N. P., et al. (2023). In-datacenter performance analysis of TPU v4. Google Research. https://doi.org/10.1145/3579371.3589357

[36] Li, T., Chen, Y., Zhang, X., & Wang, Y. (2023). Carbon-aware workload scheduling in cloud computing. IEEE Transactions on Sustainable Computing, 8(1), 12-25. https://doi.org/10.1109/TSUSC.2022.3225678

[37] Masanet, E., Shehabi, A., Lei, N., Smith, S., & Koomey, J. (2020). Recalibrating global data center energy-use estimates. Science, 367(6481), 984-986. https://doi.org/10.1126/science.aba3758

[38] McKinsey & Company. (2023). Edge computing's sustainability dividend. https://www.mckinsey.com/industries/technology/our-insights

[39] Microsoft Research. (2022). AI for energy-efficient cloud computing. https://www.microsoft.com/en-us/research/project/ai-for-cloud-efficiency/

[40] Microsoft. (2021). Water-positive cloud infrastructure: Arizona case study. https://query.prod.cms.rt.microsoft.com/cms/api/am/binary/RE4MNKx

[41] Microsoft. (2022). Zero-carbon data center strategy [Blog post]. https://blogs.microsoft.com/blog/2022/09/20/microsoft-cloud-sustainability-datacenter-innovation/

[42] Microsoft. (2023). Carbon negative by 2030. https://www.microsoft.com/en-us/corporate-responsibility/sustainability

[43] Preskill, J. (2023). Quantum computing and the entanglement frontier. Nature Physics, 19(3), 312-319. https://doi.org/10.1038/s41567-022-01820-8

[44] Shi, W., et al. (2023). Edge computing energy efficiency models. IEEE Internet of Things Journal, 10(5), 4123-4135. https://doi.org/10.1109/JIOT.2023.3242285

[45] Sze, V., Chen, Y., Yang, T., & Emer, J. (2023). Efficient processing of deep neural networks. Morgan & Claypool. https://doi.org/10.2200/S00962ED2V01Y202201CAC053

[46] The Green Grid. (2020). PUE™: A comprehensive examination of the metric (White Paper #49). https://www.thegreengrid.org/en/resources/library-and-tools/24-PUE

[47] The Green Grid. (2023). PUE: Industry benchmarks. https://www.thegreengrid.org/en/resources/library-and-tools/24-PUE

[48] The Shift Project. (2020). Lean ICT: Towards digital sobriety. https://theshiftproject.com/en/article/lean-ict-our-new-report/

[49] United Nations Environment Programme. (2023). E-waste and its environmental impact. https://www.unep.org/resources/report/global-e-waste-monitor-2023

[50] United Nations University & ITU. (2023). Global e-waste monitor 2023. https://globalewaste.org/

[51] Uptime Institute. (2023). Global data center survey 2023. https://uptimeinstitute.com/resources/asset/2023-data-center-industry-survey

[52] World Bank. (2023). Financing green digital infrastructure. https://www.worldbank.org/en/topic/digitaldevelopment/brief/financing-digital-infrastructure

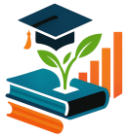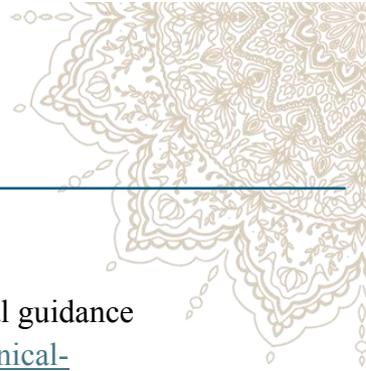

[53]   World Economic Forum. (2023). Blockchain for climate action.
https://www.weforum.org/projects/blockchain-for-climate

[54]   World Resources Institute. (2021). Greenhouse Gas Protocol: Technical guidance
for calculating scope 3 emissions. https://ghgprotocol.org/scope-3-technical-
calculation-guidance